\documentclass[%
 reprint,
showpacs,
nofootinbib,
amsmath,amssymb,
intlimits,
 prd,
]{revtex4-1}

\usepackage{graphicx}
\usepackage{dcolumn}
\usepackage{bm}
\usepackage{comment}
\usepackage{hyperref}
\usepackage{color,soul}
\usepackage{subcaption}
\def\Real{\mathbb{R}}
\def\HH{\mathcal{H}}
\def\NN{\mathcal{N}}
\def\SS{\mathcal{S}}
\def\KK{\mathcal{K}}
\DeclareMathOperator{\diag}{\mathrm{diag}}
\def\zero{^{(0)}}

\def\QQ{\mathcal{Q}}
\def\EE{\mathrm{E}}
\def\MM{\mathrm{M}}
\def\pd{\partial}

\def\surfkappa{\kappa_{(\ell)}}
\def\dd{\mathrm{d}}
\def\eps{\varepsilon}
\DeclareMathOperator{\ii}{\mathrm{i}}
\def\DD{\mathcal{D}}

\def\Lie{\pounds}

\def\JJ{\mathcal{J}}

\def\roo{\varrho}
\def\horeq{\mathbin{\dot{=}}}
\def\jE{j^{\mathrm{E}}}
\def\jM{j^{\mathrm{M}}}
\def\picirc{\pi_{\circ}}
\begin{document}

\newtheorem{theorem}{Theorem}
\newtheorem{definition}{Definition}

\preprint{APS/123-QED}

\title{The Meissner Effect for axially symmetric charged black holes}

\author{Norman G\"urlebeck}
 \email{norman.guerlebeck@zarm.uni-bremen.de}
\affiliation{%
ZARM, University of Bremen, \\
Am Fallturm, 28359 Bremen, Germany\\
DLR Institute for Space Systems\\
Linzer Str.\ 1, 28359 Bremen, Germany
}%
\author{Martin Scholtz}
\email{scholtz@utf.mff.cuni.cz}
\affiliation{
 Institute of Theoretical Physics, Charles University, V Hole\v{s}ovi\v{c}k\'ach 2, 162~00 Prague, Czech Republic
}%

\date{\today}

\begin{abstract}

  In our previous work [N.\ G\"urlebeck, M.\ Scholtz, \emph{Phys.\ Rev.\ D} {\bfseries 95} 064010 (2017)], we have shown that electric and magnetic fields are expelled
  from the horizons of extremal, stationary and axially symmetric uncharged black holes; this is called the \emph{Meissner effect} for black holes. Here, we generalize this result in several directions. First,
  we allow that the black hole carries charge, which requires a generalization of the definition of the Meissner effect.
  Next, we introduce the notion of \emph{almost isolated horizons}, which is weaker than the usual notion of isolated horizons, since the geometry of the former is not necessarily
  completely time-independent. Moreover, we allow the horizon to be pierced by strings, thereby violating the usual assumption on the spherical topology made in the definition of the weakly
  isolated horizon. Finally,
  we spell out in detail all assumptions entering the proof and show that the Meissner effect is an inherent property of black holes even in full non-linear theory.
\end{abstract}

\maketitle


\section{Introduction}\label{sec:intro}

The Meissner effect of black holes describes their property to expel any exterior magnetic field in case the black holes become extremal. This was discovered in \cite{Wald1974,Bicak-Dvorak-1976,King-1975} and was since then discussed for electromagnetic test fields \cite{Bicak-Dvorak-1980,Bicak-Janis-1985} and particular classes of black holes surrounded by a strong magnetic field \cite{Karas-Vokrouhlicky-1991,Karas-Budinova-2000,Gibbons-2014,Bicak-Hejda-2015,Bicak-Ledvinka}.

In the recent paper \cite{Guerlebeck-Scholtz-2017}, we proved that the Meissner effect holds even for \emph{generic} uncharged black holes in equilibrium that are distorted by exterior matter and electromagnetic fields. This implies that the expulsion of the electromagnetic field is not due to the specific geometries like the Kerr geometry investigated so far. Still, its physical origin is not entirely understood. Therefore, the current paper serves two purposes: First, we will show that \emph{charged} black holes exhibit the Meissner effect for strong electromagnetic fields as well, which will include all previous results. Second, we will relax as many assumptions as possible to still be able to prove the result to identify the physically necessary ones.
 
In order to understand the conceptual problems related to the first task, it is instructive to discuss first the classical example of superconductors in external magnetic fields, which lends its name to this property of black holes.

Let a neutral superconducting sphere that is above its critical temperature $T_{\mathrm{c}}$ be embedded in a magnetic field. When cooling it below $T_{\mathrm{c}}$, the magnetic field is expelled. The behavior of uncharged black holes is analogous to this standard case. Now, consider a superconducting sphere at a temperature above $T_{\mathrm{c}}$ that carries some net charge, say, because it is doted with ions. If that sphere rotates it generates itself already a magnetic field that penetrates its surface. If we apply now an external field and cool the superconductor below its critical temperature, the external field will still be expelled. However, the magnetic field produced by the moving interior charges causes still a non-vanishing magnetic flux across its surface. Hence, the Meissner effect does not predict that the total magnetic flux vanishes but only that the flux caused by external fields vanishes.

Similarly, consider the Kerr-Newman metric, which has an electric charge $Q^{\EE}$. Due to its rotation with respect to inertial observers at infinity, it also exhibits a magnetic field penetrating the horizon. This would be additional to any other magnetic test fields generated by external sources. Clearly, also here the Meissner effect discussed in \cite{Bicak-Dvorak-1980} can only make a statement about the flux caused by the external matter.

In the test field approach, both contributions to the magnetic flux, the one from the black hole and the one from the test field, can be trivially disentangled and the Meissner effect can be formulated for the test field alone, see, e.g.\ \cite{Bicak-Dvorak-1980}. However, for strong magnetic fields this disentanglement is more involved because of the non-linearity of the theory. Thus, it is not obvious, which part of the flux needs to vanish. Technically, this means that in our proof in \cite{Guerlebeck-Scholtz-2017} the argument that the vanishing charge of the horizon implies that the integration constants and consequently the fields vanish, does not apply anymore. But as described above, it should not be expected that the total magnetic flux is vanishing. It should rather be understood that the magnetic field penetrating an extremal horizon does solely depend on the properties of the horizon like its electric and magnetic charge and not on the configuration of the external matter. This can be made more precise using the initial value problem of the underlying partial differential equations. To solve them, initial data needs to be provided at the horizon and a null surface $\NN$ intersecting it. The former describes horizon properties whereas the latter the external matter and fields. The flux is now supposed to be independent of any initial data given on $\NN$. This implies then that the matter and electromagnetic fields in the exterior can be distributed arbitrarily without changing the magnetic flux across the horizon.

In contrast to our previous work \cite{Guerlebeck-Scholtz-2017}, we will allow for more general situations, in addition to allowing for a charge of the black hole. For example, we will allow strings piercing the horizon as it is the case for the C-metric \cite{Griffiths2009}, where the Meissner effect was already observed for test fields \cite{Kofron-2016}. Additionally, we will relax the equilibrium condition further generalizing the notion of isolated horizons to \emph{almost} isolated horizons. Lastly, we will not assume that the full Einstein equations hold but rather only one of its projections to allow for modified theories of gravity. 

The paper is structured as follows. In Sec.\ \ref{sec:BH-in-EQ}, we introduce the notion of charged almost isolated horizons as well as the constraint equations, which need to be solved at the horizon. In Sec.\ \ref{sec:MeissnerEffect}, we assume additionally axial symmetry to solve the constraint equation explicitly. This shows that the electromagnetic flux across the horizon depends only on horizon properties, which we parametrize in terms of the electric and magnetic charge, dipole and quadrupole moment of the almost isolated horizon, for illustration. The results are summarized and discussed in Sec.\ \ref{sec:result}, where we also give the main theorem \ref{th:ME_general}.

Throughout the paper, we use geometric units, in which $c=G=1$, and the metric signature $\diag(+1,-1,-1,-1)$. Moreover, we use the abstract index notation, cf.\ \cite{Wald-GR}. By some abuse of notation, we use the same alphabet for the abstract indices of quantities regardless of the manifolds, on which they are defined.

\section{Black holes in equilibrium}
\label{sec:BH-in-EQ}

In our previous proof \cite{Guerlebeck-Scholtz-2017}, the black holes in equilibrium were modeled by weakly isolated horizons, cf.\ \cite{Ashtekar-2002}. In order to shed some light on the origin of the Meissner effect, we will carefully spell out all assumptions, which are necessary for our proof. This allows for some generalizations of the notion of weakly isolated horizons including the possibility that it is pierced by strings, i.e., conical singularities. Thus, we give here a detailed definition of these generalized weakly isolated horizons, where we indicate deviations from the standard formalism.


Additionally, we will relax the notion of stationarity used in \cite{Guerlebeck-Scholtz-2017}, which implied essentially that all quantities on the horizon are time-independent. Here, we require this
only for the shear of the ingoing null congruence as well as one component of the electromagnetic field. This makes the black not anymore an isolated horizon but -- what we call -- an \emph{almost isolated horizon}, cf.\ Def.\ \ref{def:AIH}. 

\subsection{Non-expanding horizons}

Let $M$ be a manifold equipped with a metric $g_{ab}$ and a connection $\nabla_a$, which is compatible with $g_{ab}$. First, we introduce the notion of non-expanding horizons similar to \cite{Ashtekar-2002} in order to describe black holes in equilibrium:

\begin{definition}
  \label{def:NEH}
  A \emph{non-expanding horizon} $\HH\subset M$ is a null hypersurface with the topology $\KK \times \Real$, where $\KK$ is a compact 2-dimensional manifold such that every normal $\ell^a$ has vanishing expansion and the Ricci tensor $R_{ab}$ satisfies the following energy condition:
\begin{align}\label{eq:energy-cond}
  R_{ab}\,\ell^b \quad \text{is a causal and future-pointing vector.}
\end{align}

\end{definition}

In \cite{Ashtekar-2002}, the manifold $\KK$ was assumed to be a topological 2-sphere, which is indeed the most typical case. In the present work, though, we allow
topological defects as we discuss in Sec.\ \ref{sec:axial-symm}. Moreover, it is known that black holes distorted by external matter can even have the topology of a torus \cite{Geroch1982}. In order
to include these cases in the formalism, we keep $\KK$ general for the moment, although we will exclude the tori later. The topological defects or the choice of $\KK$ does
not affect the subsequent discussion. 

Moreover, the condition (\ref{eq:energy-cond}) was defined in terms of the energy-momentum tensor in \cite{Ashtekar-2002}. In order to keep the proof as general as possible, we do not impose
the full Einstein equations but only their projection onto the normal $\ell^a$ restricted to the horizon:
\begin{align}
  \label{eq:Einstein-00}
  \Phi_{00} \horeq \phi_0\,\bar{\phi}_0,
\end{align}
see Appendix \ref{app:np-formalism} for the standard notation of the Newman-Penrose (NP) formalism. We use the relation $\horeq$ to indicate equality of two quantities on the horizon $\HH$. For Einstein's theory, the condition (\ref{eq:energy-cond}) is equivalent to the standard definition using the energy-momentum tensor but the form \eqref{eq:energy-cond} is applicable without assuming specific field equations. Note
that Eq.\ (\ref{eq:Einstein-00}) does not exclude the presence of other types of matter $T^{\mathrm{M}}_{ab}$, we merely exclude the flux of this matter through the horizon. Additionally, it is included
that Einstein's equations are modified provided that Eq.\ (\ref{eq:Einstein-00}) still holds. 

The normal $\ell^a$ in Def.\ \ref{def:NEH} is unique up to a scaling by an arbitrary function. Let us choose any such normal and complete it to a NP null tetrad $(\ell^a, n^a, m^a, \bar{m}^a)$. Since $\ell^a$ is by
definition hypersurface orthogonal on $\HH$, it satisfies \cite{Wald-GR}
\begin{align}
  \ell_{[a} \nabla_b \ell_{c]} &\horeq 0.
\end{align}
Projecting this equation onto the aforementioned null tetrad, we get two conditions,
\begin{align}\label{eq:kappa-rhobar}
  \kappa &\horeq 0, & \roo &\horeq \bar{\roo}.
\end{align}
The former condition shows that $\ell^a$ is geodesic while the latter states that $\ell^a$ is twist-free. Notice that the conditions (\ref{eq:kappa-rhobar}) are independent of the choice of $n^a,~m^a$ and $\bar{m}^a$. The expansion of the geodesic congruence $\ell^a$ is then given by the real part of the spin coefficient $\roo$ and by Def.\ \ref{def:NEH} vanishes on the horizon. Hence, together
with Eq.\ (\ref{eq:kappa-rhobar}), we have
\begin{align}
  \label{eq:rho-0}
  \roo \horeq 0.
\end{align}
The Ricci identity (\ref{np:RI:Drho}) then implies
\begin{align}
  |\sigma|^2 + \Phi_{00} &\horeq 0.
\end{align}
Using the definition (\ref{RicciComps}) and the energy condition (\ref{eq:energy-cond}), we find
\begin{align}
  \label{eq:sigma-Phi00}
  \sigma &\horeq 0, &
                      \Phi_{00} & \horeq 0.
\end{align}
Vanishing of $\sigma$ means that $\ell^a$ is shear-free. Using Eq.\ (\ref{eq:Einstein-00}), the second condition implies
\begin{align}
  \label{eq:phi-0}
  \phi_0 &\horeq 0.
\end{align}
This is the only place where the specific relation between the Ricci tensor and the matter enter our considerations.

Now, the Ricci identity (\ref{np:RI:Dsigma}) together with (\ref{eq:kappa-rhobar}) and (\ref{eq:sigma-Phi00}) gives
\begin{align}
  \Psi_0 &\horeq 0. 
\end{align}
As discussed in \cite{Ashtekar-2002}, the energy condition (\ref{eq:energy-cond}) implies also
\begin{align}
  \Phi_{01} \horeq \Phi_{02} \horeq 0;
\end{align}
consequently,
\begin{align}
  \Psi_1 \horeq 0
\end{align}
by the Ricci identity (\ref{np:RI:deltarho}).

\subsection{Bondi-like tetrad and coordinates}
\label{sec:tetrad-coords}
The NP tetrad was arbitrary except for the fact that $\ell^a$ was chosen normal to the horizon. Subsequently, we fix some of the gauge freedom in its choice
and introduce coordinates following \cite{Krishnan-2012}. As mentioned above, the scaling of $\ell^a$ is free. We introduce equivalence classes $[\ell^a]$ as the sets
of normals, which differ only by a constant rescaling. We fix a specific equivalence class by the condition
\begin{align}
  \label{eq:WIH-condition}
  (\Lie_\ell \DD_a \ell^b)^\star &\horeq 0,
\end{align}
where $\Lie_\ell$ is the Lie derivative along $\ell^a$ and $\DD_a$ is the intrinsic connection on $\HH$ defined by $X^a \DD_a \horeq X^a\nabla_a$ for any $X^a$ tangent to $\HH$. Finally, the $\star$
denotes the pull-back of the expression to the horizon so that $(\ell_a)^\star \horeq 0$. The condition (\ref{eq:WIH-condition}) is used in \cite{Ashtekar-2002} to define weakly isolated horizons:
\begin{definition}
  \label{def:WIH}
  A \emph{weakly isolated horizon} is a pair $(\HH, [\ell^a])$, where $\HH$ is a non-expanding horizon in the sense of Def.\ \ref{def:NEH} and any element of $[\ell^a]$ satisfies the
  condition (\ref{eq:WIH-condition}).
\end{definition}

Eq.\ (\ref{eq:WIH-condition}) implies the zeroth law of black hole thermodynamics. Notice that this is merely a gauge fixing and does not restrict the geometry. However, in \cite{Ashtekar-2002}, in
the case of axially symmetric horizons, the geometry was restricted by requiring elementary flatness, as we discuss in Sec.\ \ref{sec:axial-symm}.

We define the coordinate $v$ up to an arbitrary constant by setting (cf.\ (\ref{np:operators}))
\begin{align}
  D v & \horeq 1.
\end{align}
The slices of $\HH$ of constant $v$ will be denoted by $\KK_v$ and we fix the integration constant by choosing an arbitrary slice $\KK_0$, where we set $v = 0$. The $v$-dependence
of the NP scalars describes their evolution along the horizon. Thus, we will refer to $v$ as the time coordinate, although strictly speaking it is the advanced time and, hence, a null coordinate.

Next, we
introduce arbitrary coordinates $x^I$, $I=2,3$ on $\KK_0$ and propagate them along $\HH$ by
\begin{align}
  \label{eq:xI-drag}
  \Lie_\ell x^I &\horeq 0.
\end{align}
The specific choice of the coordinates $x^I$ will be made later and depends on the topology of $\KK_0$. 

Since the vectors $m^a$ and $\bar{m}^a$ have not been specified so far, we first require that they are tangent to $\HH$. Then, we can perform a spin transformation (\ref{np:spin}) with the parameter
\begin{align}
  \chi &\horeq \frac{\ii}{2}\int_0^v (\eps-\bar{\eps})\,\dd v
\end{align}
so that after the transformation the spin coefficient $\eps$ is real on the horizon,
\begin{align}
  \label{eq:eps-real}
  \eps & \horeq \bar{\eps}.
\end{align}
Together with the Eqs.\ (\ref{eq:rho-0}) and (\ref{eq:sigma-Phi00}), which are not affected by the spin, the condition (\ref{eq:eps-real}) implies that $m^a$ is Lie dragged along the horizon,
\begin{align}
  \label{eq:m-Lie-drag}
  \Lie_\ell m^a &\horeq 0.
\end{align}

There is still a gauge freedom in the choice of $m^a$ on the initial slice $\KK_0$. At this stage, we merely require that $m^a$ is tangent to it so that $m \horeq \xi^I(x^J)\,\pd_I$ for some
$\xi^I$. A specific choice will be made later and depends again on the topology of $\KK_0$. Since we have fixed the triad $(\ell^a, m^a, \bar{m}^a)$ on the horizon by the conditions above, we have also fixed the remaining vector of the null tetrad $n^a$ on $\HH$. 

In order to propagate the coordinates $(v, x^2, x^3)$ and the null tetrad $(\ell^a, n^a, m^a, \bar{m}^a)$ off the horizon, we extend $n^a$ geodesically by requiring (cf.\ (\ref{np:operators}))
\begin{align}
  \label{eq:Deltan}
  \Delta n^a = 0.
\end{align}
Afterwards, we propagate the remaining vectors along $n^a$, i.e.\
\begin{align}
  \label{eq:Deltal-Deltam}
  \Delta \ell^a = \Delta m^a = 0.
\end{align}
The NP definitions (\ref{np:spin-coefficients}) then imply
\begin{align}
  \gamma &= \tau = \nu = 0
\end{align}
in a sufficiently small neighborhood of the horizon. 

We denote the affine parameter of $n^a$ by $r$ and set $r \horeq 0$. Finally, we propagate the coordinates off the horizon by
\begin{align}
  \Delta v &= \Delta x^I = 0.
\end{align}
In this way, we obtain a \emph{Bondi-like coordinate system} $x^\mu = (v,r,x^2,x^3)$ and a \emph{Bondi-like tetrad} of the form
\begin{subequations}
  \begin{align}
    \ell &= \pd_v + U\,\pd_r + X^I\,\pd_I, \\
    n &= -{} \pd_r, &
                    m &= \Omega \,\pd_r + \xi^I\,\pd_I
  \end{align}
\end{subequations}
in a sufficiently small neighborhood of $\HH$. By construction, we have
\begin{align}
  U & \horeq X^I \horeq \Omega \horeq 0.
\end{align}

In addition, applying the commutators (\ref{np:commutators:Ddelta}) and (\ref{np:commutators:deltadeltabar}) to the coordinate $v$ yields
\begin{align}\label{eq:palpha-mu}
  \pi &= \alpha + \bar{\beta}, &
                                 \mu &=\bar{\mu}.
\end{align}
The latter condition implies that $n^a$ is a non-twisting, hypersurface orthogonal congruence. For an explicit construction of the Bondi-like tetrad in the Kerr-Newman spacetime, see \cite{SFG-Kerr}.

To conclude this section, let us enumerate the consequences of condition (\ref{eq:WIH-condition}), which was employed in order to fix the  null normal $\ell^a$ up to rescaling by a constant factor. In the NP formalism, Eq.\ (\ref{eq:WIH-condition}) is equivalent to
\begin{align}
  \label{eq:Dpi-Deps}
  D \pi \horeq D \eps \horeq 0.
\end{align}
Combining the Ricci identities (\ref{np:RI:Dalpha}) and (\ref{np:RI:Dbeta}) with Eq.\ (\ref{eq:palpha-mu}), we find
\begin{align}
  \label{eq:Dalpha-Dbeta-deltaepsilon}
  D \alpha & \horeq D \beta \horeq \delta \eps \horeq 0.
\end{align}
Since $\eps$ is real in the gauge introduced above, the last equation together with (\ref{eq:Dpi-Deps}) in fact implies that $\eps$ is constant throughout the horizon. The quantity $\surfkappa \horeq 2\,\eps$ measures the acceleration of the normal $\ell^a$ and, thus, its constancy is interpreted as the zeroth law of black hole thermodynamics. The value of the constant $\kappa_{(\ell)}$ depends on the
scaling of $\ell_{a}$. Still, the notion of extremal weakly isolated horizons, i.e., $\kappa_{(\ell)}=0$ is unambiguously defined. 

In what follows, we will use the superscript $\zero$ to indicate the value of a quantity on the initial slice $\KK_0$. For example, the time-independence of $\pi$, Eq.\ (\ref{eq:Dpi-Deps}), implies that the value of $\pi$ on $\HH$ is
\begin{align}
  \pi &\horeq \pi\zero.
\end{align}


\subsection{Almost isolated horizon}
\label{sec:almost-isolated-horizon}

As mentioned in Sec.\ \ref{sec:tetrad-coords}, the choice (\ref{eq:WIH-condition}) of $\ell^a$ is merely a gauge fixing and can be made for any non-expanding horizon. With this choice, $\ell^a$ is a Killing vector of the induced 3-dimensional degenerate metric on $\HH$. However, $\ell^a$ is in general not a Killing vector of the full 4-dimensional metric. In particular, even the connection $\DD_a$ is not necessarily time-independent, since the spin coefficients $\mu$ and $\lambda$ can depend on $v$.

Imposing that the full connection $\DD_a$ is time-independent,
\begin{align}
  \label{eq:IH-cond}
  [\Lie_\ell, \DD_a] &\horeq 0,
\end{align}
one arrives at the notion of \emph{isolated horizon} \cite{Ashtekar-2002}. Unlike the condition for 
weakly isolated horizon (\ref{eq:WIH-condition}), the assumption (\ref{eq:IH-cond}) is a restriction of the spacetime geometry and is equivalent to the requirement that $\ell^a$ satisfies the 4-dimensional Killing equation up to the second order in the coordinate $r$. In effect, Eq.\ (\ref{eq:IH-cond}) makes $\lambda$ and $\mu$ time-independent.

In \cite{Guerlebeck-Scholtz-2017}, we assumed the presence of a stationary Killing vector, which amounts to the requirement that the horizon under consideration is, in fact, isolated in this strong sense. In the present work, we relax this assumption by allowing one component of the connection to be time dependent and it is instructive to see what is the related physical effect allowed by this weaker assumption. 

For Einstein-Maxwell spacetimes, the scalar curvature $\Lambda = 0$ and the Bianchi identity (\ref{np:BI:DPsi2}) can be used to deduce $D \Psi_2 \horeq 0$. The Ricci identity (\ref{np:RI:Dmu}) can be
solved yielding
\begin{multline}
  \mu \horeq \mu\zero\,e^{-\surfkappa\,v} \\
  + \frac{1}{\surfkappa}\left( \eth\pi\zero + |\pi\zero|^2 + \Psi^{(0)}_2 \right)\left( 1-e^{-\surfkappa\,v} \right).\label{eq:mu-horizon} 
\end{multline}
In the extremal limit, i.e.\ vanishing surface gravity $\surfkappa $, this solution simplifies to
\begin{align}
  \label{eq:mu-extremal}
  \mu &\horeq \mu\zero  + v\left( \eth\pi\zero + |\pi\zero|^2 + \Psi^{(0)}_2 \right).
\end{align}
Recall that $\mu$, which is real for the Bondi-like tetrad, represents the expansion of the (future pointing) congruence $n^a$. Clearly, the expression (\ref{eq:mu-extremal}) changes the sign once as $v$ varies
from $-\infty$ to $\infty$. In other words, the horizon is not a trapped surface over the full range of $v$. This might also happen in the non-extremal case
but not necessarily. Setting $D \mu \horeq 0$ prevents this behavior. For our proof of the Meissner effect, however, it is not necessary to impose this condition and therefore we shall not do so. Thus, the Meissner effect will hold even if $\HH$ is not a trapped surface. 

Solving the Ricci identity (\ref{np:RI:Dlambda}), one finds
\begin{align}
  \lambda &\horeq \lambda\zero\,e^{-\surfkappa\,v} + \frac{1}{\surfkappa}\left( 
            \bar{\eth}\pi\zero + (\pi\zero)^2 \right)\left( 1-e^{-\surfkappa\,v} \right), 
                                                                                            \label{eq:lambda-horizon}
\end{align}
where $\eth$ is defined by (\ref{np:eth}) and $\pi\zero$ has the spin weight $-1$. At this point, we require that $\lambda$ is time-independent on the horizon, which implies the first constraint
we need for the proof of the Meissner effect, namely
\begin{align}
  \label{eq:pi-const}
  \bar{\eth}\pi\zero +(\pi\zero)^2 &= \surfkappa\,\lambda\zero.
\end{align}
As explained above, this is a weaker condition than (\ref{eq:IH-cond}).

To summarize the previous points:
\begin{definition}
  \label{def:AIH}
  An \emph{almost isolated horizon} $(\HH, [\ell^a], n^a)$ is a weakly isolated horizon $(\HH, [\ell^a])$ in the sense of Def.\ \ref{def:WIH}, additionally equipped with a future pointing null congruence $n^a$,
  which emanates from the horizon and satisfies:
  \begin{itemize}
  \item there exists an $\ell^a \in [\ell^a]$ such that $\ell_a\,n^a \horeq 1$;
  \item $n^a \nabla_a n^b \horeq 0$;
  \item the congruence $n^a$ is non-twisting;
  \item the shear of $n^a$ is constant along $\ell^a$.
  \end{itemize}
\end{definition}

Let us stress that among the conditions listed in the Def.\ \ref{def:AIH} only the last one is constraining the geometry, the others are gauge, as discussed above.

\subsection{Initial data}
\label{sec:initial-data}

Inspecting the Ricci and the Bianchi identities, one can identify the free data
determining the geometry of the spacetime including the near-horizon geometry. It turns out
that the initial data can be prescribed on two null hypersurfaces. One is the horizon $\HH$ and the other one is the null hypersurface $\NN$ intersecting $\HH$ in the slice $\KK_0$.
Due to the properties of non-expanding horizons, cf.\ Def.\ \ref{def:NEH}, all the data on $\HH$ is, in fact, completely determined by data on $\KK_0$, although not necessarily time-independent.
In a Bondi-like tetrad, the constant $\surfkappa$ and the functions
\begin{align}
  \label{eq:free-data-K0}
  \mu\zero, \lambda\zero, \pi\zero, \xi^{(0)2}, \xi^{(0)3}
\end{align}
can be specified freely on $\KK_0$ \cite{Racz-2007,Krishnan-2012}, independently of the matter model or Einstein's equations. The Weyl scalar $\Psi_4$ on the other hand must be prescribed on the entire null hypersurface $\NN$. 

Let us impose a specific matter model, namely the electromagnetic field obeying the Maxwell equations  (\ref{np:EM:Dphi1})--(\ref{np:EM:Dphi2}) with
electric and magnetic sources, which are only two of the four Maxwell equations (\ref{np:Maxwell}). As discussed above, we do not allow for any flux of matter across the horizon, which implies that the transversal component of the current
must vanish, $\JJ_a \,\ell^a \horeq 0$. Hence, the current $\JJ_a$ must be tangent to the horizon, which, however, implies luminal motion for the carriers of the charge. Even though such a situation seems unphysical, we still allow the current in the direction of $\ell^a$ on $\HH$ in the spirit of being as general as possible. 

Under these assumptions, the electromagnetic field is determined by the value of $\phi_1\zero$ on $\KK_0$ and by $\phi_2$ on $\NN$. The sole non-vanishing component $\JJ_a n^a$ of the current on the horizon can be prescribed arbitrarily on $\HH$. Additionally, the current $\JJ^a$ can be prescribed arbitrarily on $\NN$ as long as it satisfies the continuity equation.

Finally, the scalar curvature $\Lambda$ must be specified on $\HH \cup \NN$ but imposing Einstein's equations would reduce this freedom. Note that we constrain the matter only by requiring the energy condition (\ref{eq:energy-cond}). 

\subsection{The Electromagnetic field}

So far, we have constrained the geometry of the horizon.  We will have to constrain the time-dependence of the electromagnetic field on $\HH$ as well. 
Although we do not impose full Einstein equations as explained after Eq.\ (\ref{eq:Einstein-00}), we do impose the Maxwell equations (\ref{np:EM:Dphi1}) and (\ref{np:EM:Dphi2}), which
contain only derivatives tangent to the horizon. This does not mean that we treat the electromagnetic field as a test field. For example, $\phi_1\zero$ is part of the free data on $\KK_0$ and it
affects the off-horizon geometry via the field equations one imposes, see, e.g.\ the Bianchi identity (\ref{np:BI:DeltaPsi2}).

The Eqs.\ (\ref{np:EM:Dphi1}) and (\ref{eq:phi-0}) as well as the assumptions on the current made in Sec.\ \ref{sec:initial-data} imply
\begin{align}
  \label{eq:Dphi1}
  D \phi_1 &\horeq 0.
\end{align}
In order to make the electromagnetic field time-independent on the horizon, we have to assume additionally
\begin{align}
  \label{eq:Dphi2}
  D \phi_2 & \horeq 0. 
\end{align}
Then, the latter Maxwell equation implies
\begin{align}
  \label{eq:phi1-constraint}
  \bar{\delta}\phi_1\zero + 2\,\pi\zero\,\phi_1\zero &= \surfkappa\,\phi_2\zero
\end{align}
on $\KK_0$. In other words, the assumption that the electromagnetic field is stationary on the horizon makes $\phi_1\zero$ subject to the constraint (\ref{eq:phi1-constraint}) rather than free data. 

\section{The Meissner effect}\label{sec:MeissnerEffect}

After introducing all the aforementioned notions, we can make the formulation of the Meissner effect more precise. First, the Meissner effect is a property of the magnetic flux across any part of the  horizon $\HH$, which is described by $\Im \phi_1$ at $\HH$ and has to satisfy the constraint (\ref{eq:phi1-constraint}). The Meissner effect states now that its solution is independent of
any free data, which describes exterior fields and matter, i.e., of any free data given on $\NN$ as well as $\Lambda$ and $\JJ^a$. As an additional ingredient for the Meissner effect,
we assume as it is normally done
that the horizon $\HH$ is axially symmetric and the electromagnetic field shares this symmetry. We detail its implications subsequently.

\subsection{Axial symmetry}

\label{sec:axial-symm}

Until now, all our calculations were local and independent of the topology and the geometry of the slices $\KK_v$. Those will be specified now by assuming
that $\KK_0$ is an axially symmetric 2-sphere with deficit angles at the poles. The ordinary case of topological 2-spheres treated by \cite{Ashtekar-2002} is, of course,
included by setting the deficit angles to zero.

In the axially symmetric case, one can introduce a preferred orthogonal coordinate system\footnote{Notice that the construction is similar to the construction of canonical Weyl coordinates, see, e.g., \cite{Guerlebeck-2014}.} $x^I=(\zeta, \phi)$ adapted to the axial symmetry \cite{Ashtekar2004}, in which the 2-metric reads\footnote{Here we denote the usual real constant $\pi$ by $\picirc$ in order to avoid confusion with the spin coefficient $\pi$.}
\begin{align}
  \label{eq:2-metric}
  q\zero_{IJ}\,\dd x^I\,\dd x^J &= -\frac{A}{4\picirc} \left( f^{-1}(\zeta)\,\dd \zeta^2 + f(\zeta)\,\dd \phi^2 \right),
\end{align}
where $\zeta \in [-1,1]$, $\phi\in[0,2\picirc)$, and $A$ is
the area of the slice $\KK_0$. The Killing vector associated with the axial symmetry is
\begin{align}
  \eta^a = \left(\frac{\pd}{\pd \phi}\right)^a
\end{align}
and
\begin{align}
  f(\zeta) = - \frac{4\picirc}{A} \,\eta_a\,\eta^a.
\end{align}
The function $f$ vanishes only at the north (south) pole $\zeta = +1$ ($\zeta = -1$).

In \cite{Ashtekar2004}, an additional geometric restriction is imposed, namely that $f'(\pm 1) = \mp 2$, where prime denotes the derivative with respect to $\zeta$. This amounts to
assuming elementary flatness at the poles \cite{Stephani2009}.
In the present work, we relax this assumption and allow for more general spacetimes, where the horizon might be pierced by struts. We parametrize the struts by introducing deficit angles $\alpha_{\pm}$ via
\begin{align}
  \label{eq:def-angles}
  f'(\pm 1) &= \mp\left(2 + \frac{\alpha_\pm}{\picirc}\right).
\end{align}
Nonetheless, the construction leading to the metric (\ref{eq:2-metric}) given in \cite{Ashtekar2004} is not affected by this generalization. The presence of the deficit angles is also the reason
why we parametrize the metric (\ref{eq:2-metric}) by the area $A$ of $\KK_0$ rather than by the ``radius'' $R = \sqrt{\frac{A}{4\picirc}}$ as in \cite{Ashtekar2004}. For later reference, let us denote
such $\KK_0$ by the symbol $\SS^{\alpha_+}_{\alpha_-}$. Since $\ell^a$ is a Killing vector of the metric on $\HH$, the topology of the slices $\KK_v$ cannot change in time.

In these coordinates, a convenient choice of the null vector tangent to $\KK_0$ is
\begin{align}
  m^a &= \sqrt{\frac{2\picirc}{A}}\left( \sqrt{f}\,\left(\frac{\pd}{\pd \zeta}\right)^a + \frac{\ii}{\sqrt{f}}\left(\frac{\pd}{\pd_\phi}\right)^a \right).
\end{align}
The intrinsic connection on $\KK_0$ is then fully characterized by the spin coefficient
\begin{align}
  a\zero &= \alpha\zero - \bar{\beta}\zero= m^a \bar{\delta}\bar{m}_a = - \sqrt{ \frac{\picirc}{2\,A\,f} }\,f'.
\end{align}

\subsection{Constraints}

With the restrictions introduced above, we found that the quantities $\phi_1\zero$ and $\pi\zero$ are constrained by Eqs.\ (\ref{eq:phi1-constraint}) and  (\ref{eq:pi-const}). In the extremal case
$\surfkappa = 0$, they simplify to
\begin{subequations}
  \begin{align}
    \bar{\eth}\pi\zero +(\pi\zero)^2 & = 0, \\
    \bar{\delta}\phi_1\zero + 2\,\pi\zero\,\phi_1\zero & = 0.
  \end{align}
\end{subequations}
Using the coordinates introduced in Sec.\ \ref{sec:axial-symm}, these equations read
\begin{subequations}
  \begin{align}
    (\pi\zero)' - \frac{1}{2\,f}\,\pi\zero\,f' +\sqrt{\frac{A}{2\,\picirc\,f}}\,(\pi\zero)^2 &= 0, \\
    (\phi_1\zero)' + \sqrt{\frac{2\,A}{\picirc\,f}}\,\pi\zero\,\phi_1\zero &= 0.
  \end{align}
\end{subequations}
Solving the equation for $\pi\zero$, we find
\begin{align}
  \label{eq:pi-sol}
  \pi\zero &= \sqrt{\frac{2\,\picirc}{A}}\,\frac{\sqrt{f}}{c_\pi+\zeta}, 
\end{align}
where $c_\pi$ is a complex integration constant. We discuss its possible values below. The solution for $\phi_1\zero$ is then
\begin{align}
  \label{eq:phi1-sol}
  \phi_1\zero & = \frac{c_\phi}{(c_\pi + \zeta)^2},
\end{align}
where $c_\phi$ is again a complex integration constant.

Inspection of the Ricci and the Bianchi identities shows that the integration constants $c_\phi$ and $c_\pi$ are not constrained. We will show explicitly
that they can be expressed in terms of properties of the black hole, e.g.\ its charge $\QQ_0$, and do neither depend on the data on $\NN$ nor on $\Lambda$ and $\JJ_a$.

Interestingly, the electromagnetic flux density $\phi_1\zero$ given in Eq.\ \eqref{eq:phi1-sol} is additionally independent of the concrete geometry of $\KK_0$ namely the function $f$, which is
constrained further under our assumptions. Thus, the horizon geometry is not unique but the electromagnetic flux density depends only on the four real constants encoded
in $c_\phi$ and $c_\pi$ for any geometry on $\KK_0$.

The electric and magnetic charges $Q^{\EE}$ and $Q^{\MM}$ are given by
\begin{align}
\label{eq:QQ0Def}
  \QQ_0 \equiv  Q^{\EE} + \ii\,Q^{\MM} = \oint_{\KK_0} \phi_1\zero\,\dd S,
\end{align}
where the area element of the metric (\ref{eq:2-metric}) is
\begin{align}
  \dd S = \frac{A}{4\picirc}\, \dd \zeta \wedge \dd \phi.
\end{align}
Integration yields
\begin{align}
	\label{eq:QQ0Sol}
  \QQ_0 &= \frac{A\,c_\phi}{c_\pi^2 - 1}
\end{align}
so that the value of the complex constant $c_\phi$ is fixed by the value of the charge $\QQ_0$ via
\begin{align}
  \label{eq:cphi}
  c_\phi &= \frac{\QQ_0}{A}(c_\pi^2 - 1).
\end{align}
Eq.\ \eqref{eq:QQ0Sol} shows that $c_\pi\neq\pm 1$ in order to have a well-defined charge. For $c_\pi\in(-1,-1)$, the integral Eq.\ \eqref{eq:QQ0Def} needs to be interpreted as principal value.
If we demand that the flux across any part of $\KK_0$ be well-defined, we have to assume $c_\pi \notin [-1, 1]$. 

In the case of uncharged black holes, $\QQ_0=0$, Eq.\ (\ref{eq:cphi}) immediately implies $c_\phi = 0$ and, therefore, $\phi_1\zero = 0$ by Eq.\ (\ref{eq:phi1-sol}). In this case, thus,
the Meissner effect states that both electric and magnetic field lines are expelled from the extremal horizon, as the flux of the field vanishes identically. If, on the other hand, the black hole
is charged, there is a non-vanishing flux of the field across the horizon. Then, the statement of the Meissner effect is that this flux is only due to the properties of the black hole
and it is not affected by the configuration of the matter outside the horizon. Clearly, the charge $\QQ_0$ fixes the integration constant $c_\phi$ via Eq.\ (\ref{eq:cphi}) but 
the integration constant $c_\pi$, which affects the flux in the charged case, is still not determined. However, this constant is fixed completely by quantities given on the horizon. In order to
show this explicitly, we consider the electromagnetic multipole moments associated with the horizon in the sense of \cite{Ashtekar2004}.

Although the multipole moments introduced in \cite{Ashtekar2004} are defined for $\KK_0$ being a topological 2-sphere, it is straightforward to see that the definition is applicable for our case $\KK_0 \cong \SS^{\alpha_+}_{\alpha_-}$ as well. Note that in \cite{Ashtekar2004}, they parametrize the metric (\ref{eq:2-metric}) by the function $f$ and the ``radius'' $R^2 = \frac{A}{4\picirc}$ although the latter
does not have direct geometrical meaning of a radius, as the authors also point out. In our treatment, the area $A$ and the deficit angles $\alpha_\pm$ are independent parameters of the 2-metric $q\zero_{ab}$. With this, the definition of the spherical harmonics, including the normalization and orthogonality, remains unaffected. Additionally, the coordinate $\zeta$ is still invariantly defined.

The higher complex electromagnetic moments are defined, in analogy with their flat spacetime versions, by (see\footnote{In our conventions of units, the multipole moments are defined with an additional
factor $4\picirc$.} \cite{Ashtekar2004})
\begin{align}
  \label{eq:EM-moments}
  \QQ_n &= \sqrt{\frac{4\picirc}{2\,n+1}}\,\left( \frac{A}{4\picirc}\right)^{\frac{n}{2}} \oint_{\KK_0} \phi_1\zero\,Y_{n,0}\,\dd S.
\end{align}
For convenience, we define the multipoles as complex quantities but, in fact, the quasi-locally defined electric and magnetic multipoles are given by the real and imaginary parts of $\QQ_n$, respectively. The spherical
harmonics are defined by
\begin{align}
  Y_{n,0} &= P_n(\zeta), 
\end{align}
where $P_n$ are standard Legendre polynomials of the first kind. These moments do not, in general, coincide with the asymptotic multipole moments even in the case
when there is no exterior matter as was shown in \cite{Ashtekar2004} for the quasi-local gravitational multipole moments. 

Calculating the electromagnetic dipole and quadrupole moments, we find
\begin{subequations}\label{eq:Q1-Q2}
  \begin{align}
    \QQ_1 & = -\QQ_0\, \sqrt{\frac{A}{4\,\picirc}}\, \left( c_\pi + \frac{1}{2}(c_\pi^2 -1)\log \frac{c_\pi - 1}{c_\pi + 1} \right), \\
    \QQ_2 &= \QQ_0\,\frac{A}{4\,\picirc} \left( 3 \,c_\pi^2 - 2 + \frac{3}{2}\,c_\pi(c_\pi^2-1) \log\frac{c_\pi-1}{c_\pi+1} \right).
  \end{align}
\end{subequations}
Although these quantities are formally complex, they originate from real integrals and, thus, the last terms must be interpreted as the principal values of the complex logarithms. 

In principle, only one of the moments $\QQ_1$ and $\QQ_2$ would suffice to express $c_\pi$ but in order to to this explicitly, we use an appropriate combination of the two moments so that $c_\pi$ turns out to be
\begin{align}
  \label{eq:cpi}
  c_\pi &= - \frac{2}{3\,\QQ_1}\sqrt{\frac{A}{4\,\picirc}}\left(\QQ_0 + \frac{2\,\picirc\,\QQ_2}{A} \right). 
\end{align}

As we argued above, the constant $c_\pi$ is not constrained by any data determining the exterior geometry and matter. Eq.\ (\ref{eq:cpi}) now shows that it can be in
fact related to the electromagnetic dipole and quadrupole moments, therefore, it describes the intrinsic properties of the almost isolated horizon. 

\section{Discussion}
\label{sec:result}

The purpose of the present paper was twofold. First, we showed that the Meissner effect is an inherent property of \emph{generic} black holes including distorted ones and black holes
in a broader class of theories of gravity than general relativity. Second, we elucidated all the assumptions necessary to prove the Meissner effect, which might serve as a starting point of understanding its physical origin. Because of the former, we employ the notion of weakly isolated horizon. Since the Meissner effect requires certain degree of
time independence of the horizon geometry, we introduced the notion of \emph{almost isolated horizons}, Def.\ \ref{def:AIH}, that generalizes the concept of isolated horizons. This enabled us to prove the following theorem:

\begin{theorem}\label{th:ME_general}
  Let $(\HH, [\ell^a], n^a)$ be an axially symmetric, extremal, almost isolated horizon with the topology $\HH \cong \Real \times \SS^{\alpha_+}_{\alpha_-}$ equipped with a Bondi-like tetrad $(\ell^a,n^a,m^a,{\bar m}^a)$. Let the projection (\ref{eq:Einstein-00}) of Einstein's equations and Maxwell equations (\ref{np:EM:Dphi1}) and (\ref{np:EM:Dphi2}) be satisfied with the current $\JJ^a$ proportional to $\ell^a$. In addition, let $\phi_2$ be constant along $\ell^a$ on $\HH$ and let $\phi_1$ share the axial symmetry of the metric. Then the electric and magnetic flux density $\phi_1\zero$ does not depend
  on the exterior matter described by the initial data given on $\NN$, the non-vanishing component of the current $\JJ_a$ and on the scalar curvature $\Lambda$. 
\end{theorem}

Let us briefly discuss this theorem. The physical interpretation of the topology $\SS^{\alpha_+}_{\alpha_-}$ of the slices of the horizon is that
the horizon might be pierced by struts producing topological defects at the poles. An example is the C-metric \cite{Griffiths2009}, which describes black holes accelerated by the
tension of the struts. Moreover, the treatment in \cite{Kofron-2016} shows that
in the C-metric electromagnetic test fields can exhibit the Meissner effect. This result is included here. But we go also beyond that by generalizing it to the strong field regime and by allowing possible deformations of the black hole due to external matter.

The time independence of $\phi_2$ on the horizon assumed in the theorem would be always satisfied if the electromagnetic field was stationary. This is to be expected by the analogy with
the classical Meissner effect for superconductors, where the magnetic field needs to be stationary with respect to the superconductor. 

We stress that the choice of the Bondi-like tetrad included in the theorem is just a choice of the gauge and does not constrain the geometry, cf.\ Sec.\ \ref{sec:tetrad-coords}. In this tetrad, $\phi_1\zero$ can be regarded as the flux of the electromagnetic field across the horizon. Note that the theorem includes the Meissner effect for the electric \emph{and} magnetic fields simultaneously. This does not come as a surprise since the field equations are symmetric with regard to electric and magnetic fields and the (non-)existence of electric or magnetic charges is not required.

Finally, we have shown that the electromagnetic flux density $\phi_1\zero$ is determined solely by five intrinsic properties of the black
hole, namely by the area, electric and magnetic charges and dipole moments. Hence, the matter outside the black hole, which is described by the data on $\NN$, the scalar curvature and the electromagnetic currents, does not affect $\phi_1\zero$. That is the essence of the Meissner effect. 

Additionally, we have found that $\phi_1\zero$ is independent of the metric on $\KK_0$, cf.\ (\ref{eq:phi1-sol}), and proportional to the total charge. As a consequence, we get the following
theorem, which generalizes our previous results \cite{Guerlebeck-Scholtz-2017} even in the case of uncharged black hole.

\begin{theorem}
  Under the assumptions of Theorem \ref{th:ME_general}, the electric and magnetic flux density $\phi_1$ vanishes if the almost isolated horizon is uncharged.
\end{theorem}

To conclude the paper, we discuss an open question regarding the necessity of axial symmetry raised already in \cite{Bicak-Janis-1985}. Our proof suggests that the stationarity (or at least the parts of the stationarity, we impose) is a crucial assumption and cannot be further relaxed. On the other hand, we impose the axial symmetry mainly in order to obtain the explicit solution
of the constraint equations. It seems feasible that this assumption can be relaxed entirely, which will be a subject of future investigations.

\section*{Acknowledgment}

NG acknowledges the hospitality of the Institute of Theoretical Physics, Charles University in Prague. MS is grateful for the hospitality of the ZARM, University of Bremen, and
for the financial support by the grant no.\ GAČR 17-16260Y by Czech Science Foundation. Both authors thank D.\ Giulini, J.\ Bi\v{c}\'ak and D.\ Kofro\v{n} for useful discussions.

\appendix
\section{Newman-Penrose formalism}
\label{app:np-formalism}

In this appendix, we review the definitions and equations of the Newman-Penrose (NP) formalism relevant for our discussion. We follow the
conventions of \cite{PenroseRindlerI,Stewart-1993,BST1} and we refer the reader to these publications for the full treatment of the NP formalism.

The \emph{null tetrad} consists of four null vectors $\ell^a,n^a,m^a$ and $\bar{m}^a$ normalized by the conditions $\ell^a n_a = - m^a \bar{m}_a = 1$,
while all remaining contractions vanish. The spin coefficients, which encode the connection, are given by the relations
\begin{subequations}
  \begin{align}
    \kappa &= m^a D \ell_a, &
                              \tau &= m^a \Delta \ell_a,  &
                                                            \sigma &= m^a \delta \ell_a, & \\
    \rho& = m^a \bar{\delta} \ell_a, &  
                                       \pi &= n^a D \bar{m}_a, &
                                                                 \nu &= n^a \Delta \bar{m}_a,  \\
    \lambda &= n^a \bar{\delta} \bar{m}_a , &
                                              \mu &= n^a \delta \bar{m}_a , &
  \end{align}
  \begin{align}
    \eps &= \frac{1}{2}\left[ n^a D \ell_a - \bar{m}^a D m_a\right], &  
    \beta &= 
            \frac{1}{2}\left[ n^a \delta \ell_a - \bar{m}^a \delta m_a\right] ,  \\
    \gamma &= \frac{1}{2}\left[ n^a \Delta \ell_a - \bar{m}^a \Delta m_a\right], & 
    \alpha &= \frac{1}{2}\left[ n^a \bar{\delta} \ell_a - \bar{m}^a \bar{\delta} m_a\right], 
\end{align}
\label{np:spin-coefficients}
\end{subequations}
where 
\begin{align}
  D&=\ell^a\nabla_a, &
                       \Delta &= n^a \nabla_a, &
                                                 \delta &= m^a \nabla_a, &
                                                                           \bar{\delta} &= \bar{m}^a \nabla_a.
\label{np:operators}
\end{align}
Acting on scalars, the operators $D,\Delta, \delta$ obey the commutation relations:
\begin{subequations}
  \begin{align}
    [ D,\delta ]  &=  (\bar{\pi}-\bar{\alpha}-\beta)D - \kappa
                          \Delta + (\bar{\rho}-\bar{\varepsilon}+\varepsilon)\delta + \sigma
                          \bar{\delta},\label{np:commutators:Ddelta}\\
    [\Delta, D]  &=
                            (\gamma+\bar{\gamma})D +(\varepsilon + \bar{\varepsilon})\Delta -
                            (\bar{\tau}+\pi)\delta -
                            (\tau+\bar{\pi})\bar{\delta},\label{np:commutators:DeltaD} \\ 
   [ \Delta, \delta] &=
                                   \bar{\nu}D + (\bar{\alpha}+\beta -
                                   \tau)\Delta + (\gamma-\bar{\gamma}-\mu)\delta - \bar{\lambda}
                                   \bar{\delta},\label{np:commutators:Deltadelta}\\ 
    [\delta,\bar{\delta}] &=
                                               (\mu-\bar{\mu})D + (\rho-\bar{\rho})\Delta +
                                               (\bar{\alpha}-\beta)\bar{\delta} -
                                               (\alpha-\bar{\beta})\delta.\label{np:commutators:deltadeltabar}
  \end{align}
  \label{np:commutators}
\end{subequations}
The Weyl part of the Riemann tensor is encoded in the Weyl scalars
\begin{subequations}\label{psicomps2}
  \begin{align}
    \Psi_0 &= C_{abcd}\ell^a m^b \ell^c m^d, &
                                               \Psi_1 &= C_{abcd}\ell^a n^b \ell^c m^d ,\\
    \Psi_2 &= C_{abcd}\ell^a m^b\bar{m}^c n^d, &
                                                 \Psi_3 &= C_{abcd}\ell^a n^b\bar{m}^c n^d,\\
    \Psi_4 &= C_{abcd}\bar{m}^a n^b\bar{m}^cn^d,
  \end{align}
  \label{np:Weyl scalars}
\end{subequations}
\noindent where $C_{abcd}$ is the Weyl tensor. The trace-free part of the Ricci tensor is given by the Ricci scalars
\begin{subequations}\label{RicciComps}
  \begin{align}
    \Phi_{00} &= - \frac{1}{2} R_{ab} \ell^a \ell^b, &
                                                 \Phi_{01} &= - \frac{1}{2} R_{ab} \ell^a m^b  ,\\
    \Phi_{02} &= - \frac{1}{2} R_{ab} m^a m^b , &
                                                  \Phi_{12} &= - \frac{1}{2} R_{ab} n^a m^b,
  \end{align}
  \begin{align}
    \Phi_{11} &= - \frac{1}{4} R_{ab} \left(\ell^a n^b + m^a \bar{m}^b\right) , \\
    \Phi_{22} &= - \frac{1}{2} R_{ab} n^a n^b,
  \end{align}
\label{np:Ricci components}
\end{subequations}
Finally, we introduce the symbol
\begin{align}
  \Lambda = \frac{1}{24}\,R,
\end{align}
where $R$ is the scalar curvature. These NP components of the Riemann tensor are governed by the Ricci and Bianchi identities:
\begin{widetext}
\begin{subequations}
  \begin{align}
    D \rho - \bar{\delta} \kappa &=\rho ^2+\left(\epsilon +\bar{\epsilon
                                   }\right) \rho -\kappa  \left(3 \alpha +\bar{\beta }-\pi \right)-\tau
                                   \bar{\kappa }+\sigma  \bar{\sigma }+\Phi_{00},\label{np:RI:Drho}\\
    D\sigma-\delta\kappa &= (\rho+\bar{\rho}+3\eps-\bar{\eps})\sigma - 
                           (\tau-\bar{\pi}+\bar{\alpha}+3\beta)\kappa+\Psi_0,\label{np:RI:Dsigma}\\
    D\tau-\Delta\kappa &= 
                         \rho(\tau+\bar{\pi})+\sigma(\bar{\tau}+\pi)+(\eps-\bar{\eps})\tau 
                         -(3\gamma+\bar{\gamma})\kappa+\Psi_1+\Phi_{01},\label{np:RI:Dtau}\\
    D\alpha-\bar{\delta}\eps &= (\rho 
                               +\bar{\eps}-2\eps)\alpha+\beta\bar{\sigma}-\bar{\beta}\eps - \kappa \lambda - 
                               \bar{\kappa}\gamma + (\eps+\rho)\pi + \Phi_{10},\label{np:RI:Dalpha}\\
    D\beta-\delta\eps &= (\alpha+\pi)\sigma + 
                        (\bar{\rho}-\bar{\eps})\beta-(\mu+\gamma)\kappa-(\bar{\alpha}-\bar{\pi})\eps + 
                        \Psi_1,\label{np:RI:Dbeta}\\
    D\gamma-\Delta\eps &= (\tau+\bar{\pi})\alpha + (\bar{\tau}+\pi)\beta - 
                         (\eps+\bar{\eps})\gamma - (\gamma + \bar{\gamma})\eps + \tau \pi - \nu 
                         \kappa
                         + \Psi_2 - \Lambda + \Phi_{11},\label{np:RI:Dgamma}\\
    D\lambda-\bar{\delta}\pi &= (\rho - 3\eps+\bar{\eps})\lambda + 
                               \bar{\sigma}\mu + (\pi+\alpha-\bar{\beta})\pi - 
                               \nu\bar{\kappa}+\Phi_{20},\label{np:RI:Dlambda}\\
    D\mu-\delta\pi &= (\bar{\rho}-\eps-\bar{\eps})\mu+\sigma\lambda+ 
                     (\bar{\pi}-\bar{\alpha}+\beta)\pi - \nu \kappa + \Psi_2 + 2 
                     \Lambda,\label{np:RI:Dmu}\\
    D\nu-\Delta\pi &= 
                     (\pi+\bar{\tau})\mu+(\bar{\pi}+\tau)\lambda+(\gamma-\bar{\gamma})\pi - 
                     (3\eps+\bar{\eps})\nu+\Psi_3+\Phi_{21},\label{np:RI:Dnu} \\
    \Delta\lambda-\bar{\delta}\nu &= 
                                    -(\mu+\bar{\mu}+3\gamma-\bar{\gamma})\lambda+(3\alpha+\bar{\beta}+\pi-\bar{\tau}
                                    )\nu-\Psi_4,\label{np:RI:Deltalambda}\\
    \Delta\mu-\delta\nu &= 
                          -(\mu+\gamma+\bar{\gamma})\mu-\lambda\bar{\lambda}+\bar{\nu}\pi+(\bar{\alpha}
                          +3\beta-\tau)\nu-\Phi_{22},\label{np:RI:Deltamu} \\
    \Delta\beta-\delta\gamma &= (\bar{\alpha}+\beta-\tau)\gamma - \mu \tau + 
                               \sigma \nu + \eps \bar{\nu} + (\gamma-\bar{\gamma}-\mu)\beta - 
                               \alpha\bar{\lambda}-\Phi_{12},\label{np:RI:Deltabeta}\\
    \Delta\sigma-\delta\tau& = -(\mu-3\gamma+\bar{\gamma})\sigma - 
                             \bar{\lambda}\rho - (\tau + \beta - \bar{\alpha})\tau + \kappa 
                             \bar{\nu}-\Phi_{02},\label{np:RI:Deltasigma}\\
    \Delta\rho-\bar{\delta}\tau &= (\gamma+\bar{\gamma}-\bar{\mu})\rho - \sigma 
                                  \lambda + (\bar{\beta}-\alpha-\bar{\tau})\tau + \nu \kappa - \Psi_2 - 2 
                                  \Lambda,\label{np:RI:Deltarho}\\
    \Delta\alpha-\bar{\delta}\gamma &= (\rho+\eps)\nu - (\tau+\beta)\lambda + 
                                      (\bar{\gamma}-\bar{\mu})\alpha + (\bar{\beta}-\bar{\tau})\gamma - 
                                      \Psi_3,\label{np:RI:Deltaalpha}\\
    \delta\rho-\bar{\delta}\sigma &= (\bar{\alpha}+\beta)\rho - 
                                    (3\alpha-\bar{\beta})\sigma+(\rho-\bar{\rho})\tau+(\mu-\bar{\mu})\kappa -\Psi_1 
                                    + \Phi_{01},\label{np:RI:deltarho} \\
  \delta\alpha-\bar{\delta}\beta &= \mu\rho-\lambda\sigma + 
                                     \alpha\bar{\alpha}+\beta\bar{\beta}-2\alpha\beta + (\rho-\bar{\rho})\gamma + 
                                     (\mu-\bar{\mu})\eps  -  \Psi_2 + \Lambda + \Phi_{11},\label{np:RI:deltaalpha}\\
    \delta\lambda-\bar{\delta}\mu &= (\rho-\bar{\rho})\nu + (\mu-\bar{\mu})\pi 
                                    + (\alpha+\bar{\beta})\mu+(\bar{\alpha}-3\beta)\lambda-\Psi_3 + 
                                    \Phi_{21},\label{np:RI:deltalambda}
  \end{align}
\end{subequations}
\begin{subequations}
    \begin{multline}
      D\Psi_1-\bar{\delta}\Psi_0-D\Phi_{01}+\delta\Phi_{00} = (\pi - 4 \alpha) 
      \Psi_0+2(2\rho+\varepsilon)\Psi_1-3\kappa\Psi_2+2\kappa\Phi_{11}  - 
      (\bar{\pi}-2\bar{\alpha}-2\beta)\Phi_{00}
      \\-2\sigma\Phi_{10}-
      2(\bar{\rho}+\varepsilon)\Phi_{01}+\bar{\kappa}\Phi_{02},\label{np:BI:DPsi1}
    \end{multline}
    \begin{multline}
      D\Psi_2-\bar{\delta}\Psi_1+\Delta\Phi_{00}-\bar{\delta}\Phi_{01}
      +2D\Lambda =-\lambda\Psi_0 + 2 (\pi-\alpha)\Psi_1+3\rho 
      \Psi_2-2\kappa\Psi_3+2\rho\Phi_{11}+\bar{\sigma}\Phi_{02}
      \\
      + (2\gamma+2\bar{\gamma}-\bar{\mu})\Phi_{
        00}-2(\alpha+\bar{\tau})\Phi_{01}-2\tau\Phi_{10},\label{np:BI:DPsi2}
    \end{multline}
    \begin{multline}
      D\Psi_3-\bar{\delta}\Psi_2-D\Phi_{21}+\delta\Phi_{20}-2\bar{\delta}\Lambda 
      = -2\lambda \Psi_1+3\pi\Psi_2 + 2 
      (\rho-\varepsilon)\Psi_3-\kappa\Psi_4+2\mu\Phi_{10}- 2\pi\Phi_{11}\\
      -(2\beta+\bar{\pi}-2\bar{\alpha})\Phi_{20}-2(\bar{
        \rho}-\varepsilon)\Phi_{21}+\bar{\kappa}\Phi_{22},\label{np:BI:DPsi3} 
    \end{multline}
    \begin{multline}
      D\Psi_4-\bar{\delta}\Psi_3+\Delta\Phi_{20}-\bar{\delta}\Phi_{21} 
      =-3\lambda\Psi_2 
      +2(\alpha+2\pi)\Psi_3+(\rho-4\varepsilon)\Psi_4+2\nu\Phi_{10}-2\lambda\Phi_{11}
      \\
      - (2\gamma-2\bar{\gamma}+\bar{\mu})\Phi_{20}-2(\bar{\tau}
      -\alpha)\Phi_{21}+\bar{\sigma}\Phi_{22},\label{np:BI:DPsi4}
    \end{multline}
    \begin{multline}
      \Delta\Psi_0-\delta\Psi_1+D\Phi_{02}-\delta\Phi_{01}
      =(4\gamma-\mu)\Psi_0-2(2\tau+\beta)\Psi_1+3\sigma\Psi_2\\
      +(\bar{\rho}+2\varepsilon-2\bar{\varepsilon})\Phi_{02}+ 2\sigma\Phi_{11}
      -2\kappa\Phi_{12}-\bar{\lambda}\Phi_{00}+2(\bar{\pi}-\beta)\Phi_{01},
      \label{np:BI:DeltaPsi0}
    \end{multline}
    \begin{multline}
      \Delta\Psi_1-\delta\Psi_2-\Delta\Phi_{01}+\bar{\delta}\Phi_{02}-2\delta\Lambda 
      =\nu\Psi_0+2(\gamma-\mu)\Psi_1-3\tau\Psi_2+2\sigma\Psi_3\\
      -\bar{\nu}\Phi_{00}+ 
      2(\bar{\mu}-\gamma)\Phi_{01}+(2\alpha+\bar{\tau}-2\bar{\beta})\Phi_{02}
      +2\tau\Phi_{11}-2\rho\Phi_{12},
      \label{np:BI:DeltaPsi1}
    \end{multline}
    \begin{multline}
      \Delta\Psi_2-\delta\Psi_3+D\Phi_{22}-\delta\Phi_{21}
      +2\Delta\Lambda=2\nu\Psi_1-3\mu\Psi_2+2(\beta-\tau)\Psi_3+\sigma\Psi_4\\
      -2\mu\Phi_{11}-\bar{\lambda}\Phi_{20}+ 2\pi\Phi_{12}+2(\beta+\bar{\pi})\Phi_{21
      }+(\bar{\rho}-2\varepsilon-2\bar{\varepsilon})\Phi_{22},
      \label{np:BI:DeltaPsi2}
    \end{multline}
    \begin{multline}
      \Delta\Psi_3-\delta\Psi_4-\Delta\Phi_{21}+\bar{\delta}\Phi_{22}
      =3\nu\Psi_2-2(\gamma+2\mu)\Psi_3+(4\beta-\tau)\Psi_4-2\nu\Phi_{11}\\
      -\bar{\nu}\Phi_{20}+2\lambda\Phi_{12}+2(\gamma+\bar{\mu})\Phi_{21}+(\bar{\tau}
      -2\bar{\beta}-2\alpha)\Phi_{22},
      \label{np:BI:DeltaPsi3}
    \end{multline}
    \begin{multline}
      D\Phi_{11}-\delta\Phi_{10}+\Delta\Phi_{00}-\bar{\delta}\Phi_{01}+3D\Lambda = 
      (2\gamma+2\bar{\gamma}-\mu-\bar{\mu})\Phi_{00}+(\pi-2\alpha-2\bar{\tau})\Phi_{01
      }\\
      + (\bar{\pi}-2\bar{\alpha}-2\tau)\Phi_{10}+2(\rho+\bar{\rho})\Phi_{11}+\bar{
        \sigma}\Phi_{02}
      +\sigma\Phi_{20}-\bar{\kappa}\Phi_{12}-\kappa\Phi_{21},
      \label{np:BI:DPhi11}
    \end{multline}
    \begin{multline}
      D\Phi_{12}-\delta\Phi_{11}+\Delta\Phi_{01}-\bar{\delta}\Phi_{02}+3\delta\Lambda 
      = (2\gamma-\mu-2\bar{\mu})\Phi_{01}+\bar{\nu}\Phi_{00}-\bar{\lambda}\Phi_{10}\\
      + 2(\bar{\pi}-\tau)\Phi_{11}+(\pi+2\bar{\beta}-2\alpha-\bar{\tau}
      )\Phi_{02}
      +(2\rho+\bar{\rho}-2\bar{\varepsilon})\Phi_{12}+\sigma\Phi_{21}-\kappa\Phi_{22},
      \label{np:BI:DPhi12}
    \end{multline}
    \begin{multline}
      D\Phi_{22}-\delta\Phi_{21}+\Delta\Phi_{11}-\bar{\delta}\Phi_{12}+3\Delta\Lambda 
      = \nu 
      \Phi_{01}+\bar{\nu}\Phi_{10}-2(\mu+\bar{\mu})\Phi_{11}-\lambda\Phi_{02}-\bar{
        \lambda}\Phi_{20}\\
      + (2\pi-\bar{\tau}+2\bar{\beta})\Phi_{12}+(2\beta-\tau+2\bar{\pi}
      )\Phi_{21}
      + (\rho+\bar{\rho}-2\varepsilon-2\bar{\varepsilon})\Phi_{22}.
      \label{np:BI:DPhi22}
    \end{multline}    
  \end{subequations}
\end{widetext}

The NP components of the electromagnetic tensor $F_{ab}$ are defined by
\begin{subequations}
\begin{align}
  \phi_0 &=  F_{ab} \ell^a m^b  , &
                                 \phi_2 &=  F_{ab} \bar{m}^a n^b ,\\
  \phi_1 &=  \frac{1}{2} F_{ab} \left[\ell^a n^b  -  m^a \bar{m}^b\right] 
           \label{np:EM:phi components}
\end{align}
\end{subequations}
so that the Maxwell equations read
\begin{subequations}\label{np:Maxwell}
\begin{align}
  D\phi_1 - \bar{\delta}\phi_0 &=  (\pi - 2\alpha)\phi_0 + 2\rho\phi_1 - \kappa \phi_2 -\JJ_a\,\ell^a, \label{np:EM:Dphi1}\\
  D\phi_2 - \bar{\delta}\phi_1 &=  -\lambda \phi_0  + 2\pi\phi_1 + (\rho-2\varepsilon)\phi_2 - \JJ_a\,\bar{m}^a,\label{np:EM:Dphi2}\\
  \Delta\phi_0 - \delta\phi_1 &=  (2\gamma -\mu)\phi_0 - 2\tau\phi_1 + \sigma\phi_2 + \JJ_a\,m^a,\label{np:EM:Deltaphi0}\\
\Delta\phi_1 - \delta\phi_2 &=  \nu\phi_0 - 2\mu\phi_1 + (2\beta-\tau)\phi_2 + \JJ_a\,n^a,\label{np:EM:Deltaphi1}
\end{align}
\end{subequations}
where
\begin{align}
  \JJ_a &= \frac{1}{2}\left( \jE_a + \ii\,\jM_a \right)
\end{align}
and $\jE_a$ and $\jM_a$ are the electric and magnetic currents, respectively. The Ricci tensor in the Einstein-Maxwell spacetime is given by
\begin{align}
 \Phi_{mn} &= \phi_m\,\bar{\phi}_n;
 \label{np:EM:Einstein eqs}
\end{align}
these are Einstein's equations for electrovacuum spacetime. 

By the \emph{spin transformation} in the space-like plane spanned by $m^a$ and $\bar{m}^a$ with a real parameter $\chi$, we mean the transformation
\begin{align}
 \hat{\ell}^a & \mapsto \ell^a, &
 \hat{n}^a &\mapsto n^a, &
 \hat{m}^a & \mapsto e^{2\,\ii\,\chi}m^a,
 \label{np:spin}
\end{align}
under which the spin coefficient $\eps$ transforms as
\begin{align}
 \eps & \mapsto \eps + \ii D\chi. & 
\end{align}
The quantity $\eta$ is said to have a \emph{spin weight} $s$ if it transforms like $\eta \mapsto e^{2\,\ii\,s\,\chi}\eta$ under the spin. The associated spin
raising/lowering operators $\eth$ and $\bar{\eth}$ are defined by \cite{Goldberg-1967,Stewart-1993}
\begin{align}\label{np:eth}
 \eth\eta &= \delta \eta + s \left( \bar{\alpha}-\beta \right)\eta, &
 \bar{\eth}\eta &= \bar{\delta}\eta - s \left( \alpha - \bar{\beta} \right)\eta.
\end{align}

%

\end{document}